\documentclass[11pt,a4paper]{article} 
 
\usepackage{graphicx} 
\usepackage{a4} 
\usepackage{color} 
\usepackage{xspace}
\newcommand{\compass}{{\sc Compass}\xspace}
\newcommand{\hermes}{{\sc Hermes}\xspace}
\newcommand{\zeus}{{\sc Zeus}\xspace}
\newcommand{\hera}{{\sc Hera}\xspace}
\title{Why there is no crisis of the ``spin crisis"}
\author{F. Bradamante\thanks{INFN and University of Trieste, Department of Physics, 34127 Trieste, Italy}
\ and G.K. Mallot\thanks{CERN, 1211 Geneva 23, Switzerland}}
\date{16 April 2016}
\begin{document} 
%\def\gkm{\color{blue}}
%%% 
%%% \renewcommand{\baselinestretch}{0.9}
%%% \begin{center} 
\begin{titlepage}
\maketitle
%\begin{center}
%On behalf of the COMPASS Collaboration
%\end{center}
\begin{abstract}
In a recent eprint \cite{Povh:2016kvg} it is argued that the experimental
determinations of the spin-dependent structure function $g_1$ have been
done incorrectly and that a reanalysis of those data suggests that the
original motivation to argue for a ``spin crisis", namely the small 
contribution of quark spins to the nucleon spin, is invalid. In a
subsequent note \cite{Leader:2016sli} the theoretical understanding,
as it has evolved from almost 30 years of theoretical and experimental 
scrutiny, has been shortly summarised.
In this short note, arguments are presented that the line of
reasoning in Ref.~\cite{Povh:2016kvg} does not apply, at
least not for the \compass data.
\end{abstract}
\end{titlepage}

A restitution of the strongly violated Ellis--Jaffe sum rule due to
the consideration of diffractive events has been put forward in a 
recent eprint \cite{Povh:2016kvg}.
The main argument is the claim 
that in the Deeply Inelastic Scattering (DIS) cross-section a fraction $f$ of the events is 
non-perturbative, i.e.\ diffractive, which shows no spin asymmetry, and that 
the fraction $f$ is large, of order 0.3--0.4. 
Consequently the analysing power is reduced, and  one has to rescale
the results for the polarised cross-section asymmetries by multiplying 
them by a factor $1/(1-f)$. 
Rescaling by the same factor is then required also for the first
moments of the spin-dependent structure functions $g_1$ of the proton
and the neutron, defined as
$$
\Gamma_1^{p,n} = \int_0^1 g_1^{p,n}(x,Q^2) \, dx \, .
$$
As shown in Fig.~2 of Ref.~\cite{Povh:2016kvg}, when $f$ lies between 
0.3--0.4, both the \hermes and the \compass data should be rescaled by factors of
1.4--1.6, which would bring the singlet axial coupling $a_0$ (and 
consequently $\Delta\Sigma$, the contribution of the quark spins to the spin of the 
nucleon) to about $0.6$, the value originally expected from the Ellis--Jaffe 
sum rule.
This suggestion is motivated in the eprint \cite{Povh:2016kvg} by referring to 
results from the H1 \cite{Ahmed:1992qc} and the \zeus \cite{Derrick:1994dt} 
experiments at \hera, which measured the ratio between the photoproduction 
cross-section for diffractive events and the total photoproduction
cross-section. 
This ratio is measured to be 0.30--0.40. 
Their assumption is that the fraction $f$ has to be the same %of diffractive-to-total events in 
in DIS, where analyses typically require the photon virtuality $Q^2$ to be larger 
than 1~GeV$^2$.

We have two comments to these considerations:

The first comment regards the amount of diffractive events in the \compass 
DIS data. Here only interactions of the resolved photon have to be
considered, since hard diffraction events arise from point-like virtual 
photon interactions with partons from the intrinsic proton structure.
The contribution of diffractive events to the \compass inclusive and
semi-inclusive DIS (SIDIS) event samples has been studied e.g.\ in the 
context of the analysis of hadron multiplicities \cite{Adolph:2016bga}.
For the SIDIS events at least one hadron is detected in addition to the 
scattered lepton.
The main motivations of this analysis are the extraction of the hadron 
multiplicities, $p_T$ distributions and azimuthal modulations of the
unpolarized cross-section normalised to the inclusive cross-section.
Rather than relying on particular assumptions and on measurements done at 
$Q^2 = 0$,
the diffractive production of vector mesons ($\rho^0$, $\omega$, $\phi$, \ldots),
was estimated for the actual \compass kinematics.
The evaluation is based 
on two MC simulations, one using the LEPTO \cite{Ingelman:1996mq} event 
generator simulating SIDIS free of diffractive contributions, and the other 
one using the HEPGEN \cite{Sandacz:2012at} generator simulating diffractive 
$\rho^0$ and $\phi$ production, normalised to the GPD model of 
Ref.~\cite{Goloskokov:2007nt}.
Further channels, which are characterised by smaller cross-sections and more 
particles in the final state, are not taken into account. 
Besides events with the nucleon staying intact, also events with diffractive 
dissociation of the target nucleon are simulated.
The simulation of these events includes nuclear effects, i.e.\ coherent 
production and nuclear absorption as described in Refs.~\cite{Alexakhin:2007mw,
Adolph:2012ht}.
Taking into account all these effects, the $f$ values \compass obtains for the 
inclusive event sample range from 0.04 at low $x$ and $Q^2$ to 0.003 at 
high $x$ and $Q^2$ 
\cite{Adolph:2016bga},
a result which is in line with what is known in the literature on the 
amount of diffraction in DIS \cite{BaPr2002}.
Consequently the effective dilution of the virtual photon polarisation in 
the \compass measurements is 10--100 times smaller than what
has been assumed in Ref.~\cite{Povh:2016kvg} and well inside the systematic 
uncertainties.

The second comment regards the Bjorken sum rule.
If the arguments in Ref.~\cite{Povh:2016kvg} were correct and indeed the first moments 
$\Gamma_1^p$ and $\Gamma_1^n$ had to be rescaled by a factor 
$1/(1-f) \simeq 1.5$, the Bjorken sum rule as measured from the \compass data
alone would be violated by almost 4 standard deviations. 
The Bjorken sum rule was formulated already in 1956 using current algebra, 
and reformulated in QCD. 
No discrepancy with the available data on $g_1^p$, $g_1^n$ and $g_1^d$ has 
ever been reported (apart from an early measurement corrected later). 
If the rescaling suggested in Ref.~\cite{Povh:2016kvg} would be applied, %should turn out to 
a major problem would open up for QCD.

This might be the reason, why the authors of Ref.~\cite{Povh:2016kvg} claim
an under-exhaustion of the fundamental Bjorken sum rule despite its verification
to the 9\% level by \compass \cite{Adolph:2015saz}. In their eprint, they
assert that in Ref.~\cite{Adolph:2015saz} the Bjorken sum is derived
from a fit to the scaled world data. This is incorrect, the \compass
result is obtained directly from the measured \compass $g_1^\mathrm{NS}$ non-singlet
data points in the region $0.0025<x< 0.7$ corresponding to 93.8\% of the full 
first moment. The extrapolations to $x=0$ and $x=1$ amount to 3.6\% and 2.6\%, 
respectively. For these small extrapolations a fit to the \compass non-singlet
data is used. Therefore, there is no way to turn the experimental confirmation
of the Bjorken sum rule into an under-exhaustion.

In the same way the Ellis--Jaffe sums are determined
\cite{Adolph:2015saz} at $Q^2=3~\mathrm{GeV}^2$. We obtain for the proton 
$\Gamma_1^p = 0.139 \pm 0.010$ and for the neutron $\Gamma_1^n = -0.041\pm0.013$.  
The theoretical values are $0.172 \pm 0.003$ and $-0.017\pm 0.003$,
respectively.
It remains unclear why in Ref.~\cite{Povh:2016kvg} it is stated that 
``... however with the realization of the idea presented in this paper the 
Bjorken sum as well as the Ellis-Jaffe-sum rule are in accord with the 
data naturally", whereas ``this realisation" leads to a clear violation 
of the Bjorken sum rule.

\section*{Acknowledgements}
We thank our \compass colleagues for fruitful discussions and clarifications,
in particular Y.~Bedfer, M.~Faessler, F.~Kunne, E.M. Kabu\ss, A.~Kotzinian, A.~Sandacz, 
E.~Seder, M.~Stolarski, and M.~Wilfert. Useful discussions with V.~Barone are 
also acknowledged.

\end{document}